\documentclass[12pt]{article}
\usepackage[dvips, letterpaper,hmargin=1in,vmargin=1in]{geometry}

\usepackage{graphicx,amsmath,amsfonts,amssymb,aas_macros}

\usepackage{titlesec}
\titleformat{\section}{\large\bfseries}{\thesection}{1em}{}

\newcommand{\SMgroup}{\ensuremath {\mathrm{SU}}(3)_{c}\times {\mathrm{SU}}(2)_{L}\times{\mathrm U}(1)_{Y}}
\newcommand{\Uoneprime}{\ensuremath {\mathrm U}(1)_{V}}

\DeclareMathOperator{\real}{Re}
\DeclareMathOperator{\imag}{Im}

\begin{document}

\begin{titlepage}
 
\begin{flushright}
\normalsize{PI-PARTPHYS-318}
\end{flushright}

\setcounter{page}{1}

\vspace*{0.2in}

\begin{center}

\hspace*{-0.6cm}\parbox{17.5cm}{\Large \bf \begin{center}
New stellar constraints on dark photons
\end{center}}

\vspace*{0.5cm}
\normalsize

\vspace*{0.5cm}
\normalsize

{\bf Haipeng An$^{\,(a)}$, Maxim Pospelov$^{\,(a,b)}$ and Josef Pradler$^{\,(c)}$}

\smallskip
\medskip

$^{\,(a)}${\it Perimeter Institute for Theoretical Physics, Waterloo,
ON, N2L 2Y5, Canada}

$^{\,(b)}${\it Department of Physics and Astronomy, University of Victoria, \\
  Victoria, BC, V8P 1A1 Canada}

$^{\,(c)}${\it Department of Physics and Astronomy, Johns Hopkins University, \\
  Baltimore, MD, 21210 USA}

\smallskip
\end{center}
\vskip0.2in

\centerline{\large\bf Abstract}

We consider the stellar production of vector states $V$ within the
minimal model of "dark photons".  We show that when the St\"uckelberg
mass of the dark vector becomes smaller than plasma frequency, the
emission rate is dominated by the production of the longitudinal modes
of $V$, and scales as $\kappa^2 m_V^2$, where $\kappa$ and $m_V$ are
the mixing angle with the photon and the mass of the dark state. This
is in contrast with  widespread assertions in the literature that
the emission rate decouples as the forth power of the mass. We derive
ensuing constraints on the $(\kappa, m_V)$ parameter space by calculating
the cooling rates for the Sun and horizontal branch stars.  We find that
stellar bounds for $m_V < 10$~eV are significantly strengthened, to
the extent that all current "light-shining-through-wall" experiments
find themselves within deeply excluded regions.

\vfil

\end{titlepage}

\section{Introduction}

The Standard Model of particles and fields (SM) can be naturally
extended by relatively light neutral states.  Almost all possible ways
of connecting such states to the SM have been explored, and several of
such ways stand out as the most economical/natural. One of the most
attractive possibilities is the so-called "hypercharge portal", or
"kinetic mixing" portal that at low energy connects the
electromagnetic current with another massive photon-like state
\cite{Holdom:1985ag}.  This model has been under intense scrutiny in
the last few years, both experimentally and observationally.  The
interest to this model is fueled by attractive (yet speculative)
possibilities: the dark vector can be a promising mediator of the dark
matter-SM interaction~\cite{Boehm:2003hm}, or form super-weakly
interacting dark matter
itself~\cite{Pospelov:2008jk,Redondo:2008ec}. Dark vectors were
proposed as a possible solution to the muon $g-2$
discrepancy~\cite{Pospelov:2008zw}, and have been searched for (so far
with negative results), both at high energy and in medium energy
high-intensity particle physics experiments.

The region of small vector masses, $m_V <$~eV, can also be very
interesting.  On the theoretical side, there are speculations of dark
photons contributing to dark matter (via an initial condensate-like
state) \cite{Piazza:2010ye} and dark radiation \cite{Jaeckel:2008fi}.
But perhaps more importantly, there are some hopes for the terrestrial
detection of dark photons. So, far several avenues have been proposed:
one can attempt observing a "visible-dark-visible" oscillation chain in
"light-shining-through-wall" experiments (LSW) \cite{Ahlers:2007qf}.
The quanta of dark photons emitted from the Sun can be searched for
with "helioscopes" \cite{Redondo:2008aa}, neutrino-
\cite{Gninenko:2008pz} and dark matter experiments
\cite{talks,Horvat:2012yv}.  Some of these exciting possibilities have
been summarized in the recent review~\cite{Jaeckel:2010ni}.  We will
refer to all proposals and experiments aimed at detection of dark
vectors, produced astrophysically or in the laboratory, as direct
searches.

At the same time, it is well-known that for many light ($m_V < {\rm
  keV}$) and weakly-coupled exotic particles the astrophysical
constraints are often far stronger than direct laboratory
constraints~\cite{Raffelt:1996wa}. The astrophysical constraints are
very important for the dark vectors as well, as they determine a
surviving fraction of the parameter space that can be explored in
direct searches. The most important limits to recon with are the
constraints on the emission of dark vectors from solar luminosity,
from the horizontal branch stars, neutron star and supernovae cooling rates.

To date, the only in-depth analysis of astrophysical bounds on sub-keV
dark vectors was performed by Redondo in \cite{Redondo:2008aa}, where the production of longitudinal modes of the dark photon is treated incorrectly. We trace the mistake traced to a
wrong use of the in-medium polarization effects for longitudinal modes.   In
this paper we re-assess these bounds, provide correct calculations for
the dark photon emission rates, and strengthen the astrophysical
bounds in the LSW region by as much as ten orders of magnitude. Our
findings significantly reduce the parameter space available for the
direct searches and affect or completely change the conclusions of
many papers written on this subject. In a separate forthcoming
publication we will address new limits imposed by the most advanced
WIMP detectors on the solar emission of dark vectors
\cite{ournextpaper}.

This paper is organized as follows. The next section introduces the
minimal model of the dark vector, and explains the main scaling of its
production rate with $m_V$. Section~3 contains technicalities of the
in-medium production of the dark vector. Section~4 contains practical
formulae for the stellar emission rates, in application to the Sun and
horizontal branch stars, and sets the constraints on the mass-mixing parameter
space.  We reach our conclusions in section~5.

\section{Dark photon production, in vacuum and in a medium}
\label{sec:2}

The minimal model of "dark vectors" extends the SM gauge group $\SMgroup$ by an
Abelian factor $\Uoneprime$. Kinetic mixing of the hypercharge field
strength $F_{\mu\nu}^Y$ with the field strength $V_{\mu\nu}$ of
$\Uoneprime$ links the SM to the new physics sector, while SM fields are
assumed to be neutral under $\Uoneprime$. We are interested in processes
far below the electroweak energy scale, for which the relevant
low-energy Lagrangian takes the form
\begin{align}
  \label{eq:L}
  \mathcal{L} = -\frac{1}{4} F_{\mu\nu}^2-\frac{1}{4} V_{\mu\nu}^2 -
  \frac{\kappa}{2} F_{\mu\nu}V^{\mu\nu} + \frac{m_V^2}{2} V_{\mu}V^{\mu}
  + e J_{\mathrm{em}}^{\mu} A_{\mu} .
\end{align}
Here $F_{\mu\nu} = \partial_{\mu} A_{\nu} -\partial_{\nu} A_{\mu} $ is
the photon field strength and $V_{\mu}$ is the ``hidden photon'' (also
known as "dark vector", "secluded vector", "dark photon" etc---an
equivalent set of names). The coupling of $A_\mu$ and $V_\mu$ is
regulated by the kinetic mixing parameter $\kappa$, redefined in an
appropriate way to absorb the dependence on weak mixing angle. For all
calculations in this paper we use $\kappa\ll 1$, and consider only
leading order terms in the mixing angle.  Finally $
J_{\mathrm{em}}^{\mu}$ is the usual electromagnetic current with
electric charge $e<0$.

It is important to comment on the origin of $m_V$ in (\ref{eq:L}). The
simplest possibility is that $m_V$ is a St\"uckelberg-type
mass. Because of the conservation of the Abelian vector current, $m_V$
remains protected against sensitivity to UV scales, and such a model is
technically natural even with very small $m_V$. An alternative
generic possibility is a new scalar field(s) charged under $\Uoneprime
$ that develops a vacuum expectation value that Higgses the hidden
group. This introduces a new interaction term of the physical hidden
Higgs with vectors, $g' m_V h' V_{\mu}^2$, as well as $h'$
self-interaction (see {\em e.g.} \cite{Batell:2009yf}). It is 
well understood that in the limit of $m_V$ and $m_{h'}$ small compared to all energy scales in the problem, the production of dark
sector states is dominated by the dark Higgsstrahlung
\cite{Pospelov:2008jk,Batell:2009yf}, or equivalently, by the
pair-production of the $\Uoneprime $-charged Higgs scalar
fields. Importantly, this process is insensitive to the actual mass of
$m_V$ in the small mass limit, and schematically
\begin{equation}
\label{rate_higgsed}
{\rm Rate}_{SM \to V+h'} \propto  \alpha' \kappa^2 (m_V)^0,
\end{equation}
where we show only the dependence on dark sector parameters, leaving the
SM part of the $V+h'$ production process completely general; $\alpha'
= (g')^2/(4 \pi)$ is the square of the coupling of dark Higgs to
$V_\mu$.  For sub-keV dark vectors and Higgses, all previously derived
constraints on "millicharged particles" apply \cite{Davidson:2000hf},
and limit the $\kappa g'$ combination to be below $\sim 10^{-13}$.  The
technical reason for not having any small $m_V$ suppression of the
rate (\ref{rate_higgsed}) despite the interaction term $g' m_V h'
V_{\mu}^2$ being proportional to $m_V$ is of course tied to the
production of the longitudinal modes of $V$ in $V+h'$ final state.

The models with the hard ({\em i.e.} St\"uckelberg) mass $m_V$ behave differently 
as the production rate of dark vectors {\em has to} decouple in the small $m_V$ limit. 
The easiest way to see this is to restrict the interaction terms in (\ref{eq:L}) to on-shell 
$V_\mu$, using $ \partial_\mu V^\mu = 0 $ and to leading order in $\kappa$, $\partial_{\mu} V^{\mu\nu}  = -m_V^2 V^\nu$, so that
\begin{equation}
{\cal L}_{\rm int} = -
  \frac{\kappa}{2} F_{\mu\nu}V^{\mu\nu} 
  + e J_{\mathrm{em}}^{\mu} A_{\mu}  ~~ \xrightarrow{\rm on-shell~V} ~~  
{\cal L}_{\rm int} = - \kappa m_V^2 A_\mu V^\mu
  + e J_{\mathrm{em}}^{\mu} A_{\mu}.
\label{reduction}
\end{equation}
This expression is of course explicitly gauge invariant under $A_\mu \to A_\mu + \partial_\mu \chi$
due to the current conservation and on-shellness of $V_\mu$ conditions:
\begin{equation}
\partial_\mu J_{\mathrm{em}}^{\mu} =0;~~~ \partial_\mu V^\mu = 0.
\end{equation}

The appearance of $m_V^2$ in the coupling of $V_\mu$ and $A_\mu$ shows
that two sectors are decoupled in $m_V=0$ limit. The most important
question in considering the production of $V_\mu$ states is the
scaling of the production rate with $m_V$, in vacuum and inside a medium.
The existing literature on the subject \cite{Redondo:2008aa} and its
subsequent follow-up papers claim that in-medium production decouples
as ${\rm Rate}_{ \,SM \to V} \propto \kappa^2 m_V^4$ in the small
$m_V$ limit.  This inference is wrong.

\begin{figure}
\centering
\includegraphics[height=1.5in]{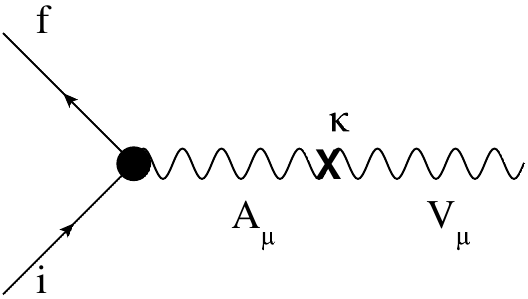}
\caption{Illustration of the dark photon emission process by the electromagnetic current.}\label{fig:feynman}
\end{figure}

To demonstrate our point we consider a generic production process
$i\to f + V$ due to (\ref{reduction}), where $i,\,f$ are any initial,
final states of the SM particles. A schematic drawing of such a
process is shown in Fig.~\ref{fig:feynman}. Without loss of generality
we assume that $V$ is emitted in $z$-direction, so that its
four-momentum $k_\mu$ is given by $(\omega, 0, 0, |\vec{k}|)$, with
$\omega^2-\vec{k}^2=m_V^2$. Moreover, we assume that the energy of
the emitted $V$ is much larger than its rest mass, $\omega\gg m_V$. Three
polarization states can be emitted: two transverse states $V_T$ with
polarization vectors $\epsilon^T = (0,1,0,0,)$ and $(0,0,1,0)$,
and one longitudinal mode $V_L$ with polarization vector
$\epsilon^L = m_V^{-1}(|\vec{k}|,0,0,\omega)$.  In all cases
$\epsilon_\mu^2 =-1$ and $\epsilon_\mu k^\mu=0$.

We include a boundary-free medium via some conducting plasma, characterized by the plasma frequency 
$\omega_p$. We consider two regimes, [almost] vacuum: $ \omega_p \ll m_V \ll \omega$, and in-medium:
$ m_V \ll \omega_p \ll \omega$. The choice of $|\vec{k}|,\omega \gg \omega_p$ is not essential, and we consider 
all ranges of $\omega$ in the next section. 
The matrix element for the production process  
induced by (\ref{reduction}) is given by 
\begin{equation}
\label{amplitude}
{\cal M}_{i\to f+V_{T(L)}} = \kappa m_V^2\, [e {J_{\rm em}}_\mu]_{fi} \,\langle  A^\mu, A^\nu  \rangle\, \epsilon_\nu^{T(L)},
\end{equation} 
where $\langle  A^\mu, A^\nu  \rangle$ stands for the photon propagator with input momentum $k_\mu$, and 
$[e J_{\rm em}^\mu]_{fi}$ is the 
matrix element of the electromagnetic current. We disregard various $m_V$-independent 
phase factors and normalizations, as our goal in this section 
is to only consistently follow the powers of $m_V$. 

For convenience, we fix the photon gauge to be Coulomb, $\nabla_i A_i=0$,
but same results can be obtained in other gauge, of course. The photon propagator for the production of the 
transverse modes is given by (see, {\em e.g.} \cite{Braaten:1993jw}):
\begin{eqnarray}
\langle A_j, A_l \rangle = \frac{\delta_{jl}^\perp}{\omega^2-|\vec{k}|^2-\omega_p^2  }
= \frac{\delta_{jl}^\perp}{m_V^2-\omega_p^2  } \longrightarrow 
\delta_{jl}^\perp\times \left\{\begin{array}{c}
m_V^{-2}  ~~ {\rm at} ~ m_V\gg \omega_p,  \\
-\omega_p^{-2} ~~ {\rm at} ~ m_V\ll \omega_p,
\end{array}\right. 
\end{eqnarray}
where $\delta_{jl}^\perp$ is the projector onto transverse modes.
Because of the existence of two different regimes for the transverse
modes of the photon, the amplitude of the $V^T$ production is
$m_V$-independent in vacuum, and $m_V^2$-suppressed in medium.

The production of the longitudinal modes in Coulomb gauge is mediated by the $ \langle A_0, A_0 \rangle$
part of the propagator. If $k\simeq \omega\gg \omega_p$, this propagator is unaffected by the medium
\cite{Braaten:1993jw}, 
\begin{equation}
\label{A0A0}
\langle A_0, A_0 \rangle = \frac{1}{|\vec{k}|^2}.
\end{equation}
The difference in the behavior of the two parts of the photon propagator
can be readily understood. While the propagating transverse modes of
the photon are close to mass shell, and plasma effects can affect them
easily no matter how large $\omega$ is, the non-propagating piece
$\langle A_0, A_0 \rangle$ cannot be changed in a causal manner if
$|\vec{k}|$ is larger than {\em e.g.} the inverse distance between
particles in the plasma. Plugging the expression (\ref{A0A0}) for the
propagator together with the explicit form of $\epsilon_\mu^L$ into
(\ref{amplitude}), we arrive at the conclusion that the amplitude for
the production of the longitudinal mode is proportional to the first power
of $m_V$.

Collecting all relevant factors in one expression, 
we get the all-important $m_V$ scalings for the production of both transverse and longitudinal 
modes:
\begin{eqnarray}
\label{scalingT}
{\rm Rate}_{\,SM\to V_T} \propto \left\{\begin{array}{c}
\kappa^2  ~~~~~~~~~~~ {\rm in~vacuum}, ~ m_V\gg \omega_p,  \\
\kappa^2 m_V^4\omega_p^{-4} ~~ {\rm in ~medium}, ~ m_V\ll \omega_p.
\end{array}
\right.
\end{eqnarray}
Performing the same estimate for the production of the longitudinal mode, (and keeping 
$\omega \gg \omega_p$) we get
\begin{equation}
\label{scalingL}
{\rm Rate}_{\,SM\to V_L} \propto  \kappa^2 m_V^2 \omega^{-2} ,~~{\rm both~in~vacuum~and~in~medium}.
\end{equation}
Notice the quadratic, and not quartic dependence on the dark photon
mass in (\ref{scalingL}), in contrast with conclusions of
Ref. \cite{Redondo:2008aa}.  The $m_V$-scaling of the production rate
for the transverse modes, Eq.~(\ref{scalingT}), is of course a
well-known result in the ``dark photon'' literature. It exhibits $m_V^4$
decoupling at small $m_V$, but as it turns out, this was incorrectly
extended to the production of the longitudinal modes.  (Thus we also learn
that ``nature does not like to skip an order,'' and compared to
the Higgsed case of an $O(m_V^0)$-rate (\ref{rate_higgsed}) the
emission of St\"uckelberg vectors occurs in lowest possible $O(m_V^2)$
order.)  We also note in passing that for some processes, only the
production of of the longitudinal modes of $V_\mu$ is actually
possible.  A well-studied process is the $K^+\to \pi^+ V$ decay
\cite{Pospelov:2008zw}, which is fobidden for the transverse modes,
but has the expected $m_V^2$ scaling at small $m_V$ for the
longitudinal modes.

The correct scaling (\ref{scalingL}) will bring about momentous change
in all estimates of stellar cooling rates.  For the favorite LSW
region with $m_V \sim 10^{-3}$~eV, one should expect that the previous
literature underestimates the solar cooling rates by as much as
$m_V^2/\omega_p^2 \sim 10^{-10}$, and correspondingly, one expects the
tightening of the constraints by the same large factor, once the
production of the longitudinal mode of the dark vector is treated
correctly. In the two subsequent sections we perform this analysis in
some detail.

In closing this section, it is  important to realize that the $m_V^2$-scaling of the ${\,SM\to V_L}$ rate
derived in Eq. (\ref{scalingL}) is not going to change if one abandons the
$\omega \gg \omega_p$ approximation. Indeed, in the Coulomb gauge the only "sources" of 
$m_V$ (at small $m_V$) in the entire problem are the coupling in the Lagrangian and the 
expression for $\epsilon^L$. Therefore, the counting of powers of $m_V$ is the same, 
and the $m_V^2$ scaling is  preserved even when 
the  medium effects significantly modify the propagator (\ref{A0A0}) at $\omega \sim \omega_p$. 
At the same time, 
these effects will affect the {\em normalization } of $m_V^2$ in (\ref{scalingL}), and in order to derive 
an accurate expression for the rate, we have to go through a more detailed account of 
in-medium effects in the next section.

\section{Plasma production rate of dark photons}
\label{sec:3}

Inside a medium, the propagation of the electromagnetic field is
determined by the electromagnetic polarization tensor $\Pi^{\mu\nu} =
e^2 \langle J_{\rm em}^\mu ,J_{\rm em}^\nu\rangle$. Due to $k_\mu
J_{\rm em}^\mu = 0$ where $k=(\omega,\vec k)$ is the four-momentum
flow inside the polarization tensor, $\Pi^{\mu\nu}$ can be
parameterized as
\begin{equation}
\Pi^{\mu\nu} = \Pi_T \sum_{i=1,2} {\epsilon^{T}}^{\mu}_i {\epsilon^T}^\nu_i + \Pi_L {\epsilon^{L}}^\mu {\epsilon^L}^\nu \ ,
\end{equation}
where $\epsilon^{T,L}$ are the transverse and longitudinal
polarization vectors of a vector boson with momentum $k$. In
particular,
\begin{equation}\label{epsilonL}
\epsilon^L = \frac{1}{\sqrt{\omega^2 - |\vec k|^2}} (|\vec k|,\omega \frac{\vec k}{|\vec k|}) \ .
\end{equation}

In the Coulomb gauge, the propagator of the electromagnetic field can be written as
\begin{eqnarray}
\langle A^i, A^j \rangle &=& \frac{1}{\omega^2 - |\vec k|^2 - \Pi_T} \left(\delta^{ij} - \frac{k^i k^j}{\vec k^2}\right) \ , \nonumber \\
\langle A^0, A^0 \rangle &=& \frac{1}{|\vec k|^2 - \frac{|\vec k|^2}{\omega^2 - |\vec k|^2} \Pi_L} \ ,
\end{eqnarray}
where $\epsilon^L_0 = |\vec k|/\sqrt{k_\mu k^\mu}$ has been used.
Notice that while the definition of $\Pi_T$ is usually uniform across
the literature, the definition of $\Pi_L$ varies, and {\em e.g.} in
Ref.~\cite{Braaten:1993jw} it is defined differently, $\Pi_L^{\rm
  ~Ref.\, {\tiny \cite{Braaten:1993jw}}} = \frac{|\vec k|^2}{\omega^2 -
  |\vec k|^2}\Pi_L^{\rm ~this~ work} $.  With the explicit expression
of the photon propagator, the matrix element for the dark photon
emission in Eq.~(\ref{amplitude}) can further be written as
\begin{eqnarray}\label{amplitude2}
{\cal M}_{i\rightarrow f+V_T} &=& - \frac{\kappa m_V^2}{m_V^2 - \Pi_T} [e J^\mu_{\rm em}]_{fi} \epsilon^T_\mu \ , \nonumber \\
{\cal M}_{i\rightarrow f+V_L} &=& \frac{\kappa m_V^2}{ m_V^2 - \Pi_L} \frac{m_V^2}{|\vec k|^2} [e J^0_{\rm em}]_{fi} \epsilon^L_0 \ .
\end{eqnarray}   
Using the condition $k_\mu J^\mu_{\rm em} = 0$ and Eq.~(\ref{epsilonL}), it is easy to show that 
\begin{equation}
J^0_{\rm em} \epsilon^L_0 = - \frac{|\vec k|^2}{m_V^2} J_{\rm em}^\mu \epsilon^L_\mu \ .
\end{equation}
Therefore, Eq.~(\ref{amplitude2}) can be further simplified to
\begin{equation}\label{amplitude3}
{\cal M}_{i\rightarrow f+V_{T,L}} = - \frac{\kappa m_V^2}{m_V^2 - \Pi_{T,L}} [e J^\mu_{\rm em}]_{fi} \epsilon^{T,L}_\mu \ .
\end{equation}
The general expression for $\real \Pi_{T}$ and $\real \Pi_{L}$ can be
found in Ref.~\cite{Braaten:1993jw}.  Thus we find that the emission
of dark vectors is given by the vacuum matrix element for the emission
of massive photons, $[e J^\mu_{\rm em}]_{fi} \epsilon^{T,L}_\mu$, with
fiducial photon mass $m_\gamma = m_V$ and multiplied by the {\em
  effective} mixing angles, defined according to,
\begin{eqnarray}\label{eq:kappa}
\kappa_{T,L}^{2} = \frac{\kappa^{2} m_{V}^{4}}{(m_{V}^{2} - \real \Pi_{T,L})^{2} + (\imag \Pi_{T,L})^{2}} \ . 
\end{eqnarray}

At finite temperature $T\neq 0$, the imaginary parts of $\Pi_{T,L}$
are related to the rates at which the respective distribution
functions approach equilibrium~\cite{Weldon:1983jn}. Detailed
balance equation allows this to be expressed exclusively in terms of the
absorption rate $\Gamma^{\rm abs}_{T(L)}$  of in-medium transverse (longitudinal) massive photons,
\begin{equation}\label{eq:impi}
\imag \Pi_{T,L} (\omega,|\vec k|) = -\omega\left(1-e^{-\omega/T}\right)\Gamma^{\rm abs}_{T,L}(\omega,|\vec k|) \ .
\end{equation}

One should note that the on-shell condition of the outgoing dark photon imposes the relation $\omega^2 = |\vec k|^2 + m_V^2$. Therefore, in this sense, $\Gamma^{\rm abs}_{T,L}$ can be understood as the absorption rate of the transverse and longitudinal mode of dark photon as if it couples to the electromagnetic current with the same coupling constant as the photon field. 

It will be convenient to express $\Gamma^{\rm abs}_{T(L)}$ in terms of a
differential production rate ${d\Gamma^{\rm prod}_{T,L}}/({d\omega dV})$ per
frequency interval and volume. To this end we write a Boltzmann-type
equation for the photon distribution function $dn_{T,L}/d\omega$,
\begin{equation}
  \frac{dn_{T,L}}{d\omega dt} = 
  \frac{d\Gamma^{\rm prod}_{T,L}}{d\omega dV}\frac{g_{T,L}}{1-e^{-\omega/T}} - \frac{dn_{T,L}}{d\omega} \Gamma^{\rm abs}_{T,L} .
\end{equation}
Using detailed balancing and noting that the equilibrium dark photon
distribution is given by $ dn_{T,L}/d\omega = g_{T,L}\omega |\vec
k|/(2\pi^2)/(e^{\omega/T} - 1)$, where $g_{T} = 2$ and $g_L = 1$, are the degeneracies for the transverse and longitudinal modes respectively, this yields,
\begin{equation}
\Gamma^{{\rm abs}}_{T,L} = \frac{2\pi^{2}}{\omega |\vec k|} \frac{d\Gamma^{{\rm prod}}_{T,L}}{d\omega dV}e^{\omega/T} \ .
\end{equation}

Therefore, according to Eqs.~(\ref{amplitude3}), (\ref{eq:kappa}), 
we conclude that the differential production rate of 
transverse and longitudinal modes of dark vectors, ${d\Gamma^{{\rm prod},V}_{T,L}}/(d\omega dV)$, can be written as
\begin{eqnarray}\label{eq:gammaV}
\frac{d\Gamma^{{\rm prod},V}_{T,L}}{d\omega dV} = \kappa_{T,L}^{2} \frac{d\Gamma^{\rm prod}_{T,L}}{d\omega dV} \ .
\label{GammaV}
\end{eqnarray}

Current conservation demands that both $\real \Pi_L$ and $\imag\Pi_L$
are proportional to $m_V^2$.  Therefore, $\kappa_L^2$ will remain
$m_V$-independent in the small $m_V$ limit as can be seen from
(\ref{eq:kappa}) when expressed as
\begin{equation}\label{eq:kappaL}
\kappa_{L}^{2} = \frac{\kappa^{2} } {\left(1 - \frac{\real \Pi_L}{m_V^2} \right)^{2} + \left(\frac{\imag \Pi_{L}}{m_{V}^{2}}\right)^{2}} \ ,
\end{equation}
From Eqs.~(\ref{eq:kappa}) and (\ref{eq:kappaL}) it is obvious that in
the region $m_V^2 \ll {\real \Pi_T}$, $\kappa_T^2$ scales as $m_V^4$,
whereas $\kappa_L^2$ scales as $m_V^0$. However, $\Gamma^{{\rm abs}}_{L}, 
\Gamma_L^{\rm prod}$ scale
as $m_V^2$ due to the current conservation. Therefore, the production
rate of the transverse modes is suppressed by $m_V^4$ in the small
$m_V$ limit whereas the production rate of the longitudinal mode is
only suppressed by $m_V^2$, in agreement with our qualitative
discussion in the previous section. To avoid confusion 
we would like to stress again that $\Gamma^{{\rm abs}}_{L}$ 
describes the damping of the longitudinal photon modes, and not 
what one would call plasmons. The damping of plasmons 
will be $m_V^2$-independent, and can be obtained from $\Gamma^{{\rm abs}}_{L}$ 
upon the appropriate rescaling. 
From Eq.~(\ref{eq:kappa}) one can see that $\Gamma^{{\rm prod},V}_{T,L}$ reaches resonance at $m_V^2 = {\rm Re}\Pi_{T,L}$, and can be written as 
\begin{equation}\label{eq:gammares}
\left.\frac{d\Gamma^{{\rm prod,}V}_{T,L}}{d\omega dV} \right|_{\rm res} = \frac{\kappa^2 m_V^4\sqrt{\omega^2 - m_V^2}}{2\pi(e^{\omega/T} - 1)} \delta(m_V^2 - {\rm Re}\Pi_{T,L}) \ ,
\end{equation} 
which is independent of the details of the production processes. 

The resonant production can be  understood as a thermal bath of photons which slowly transits into dark photons. The transition amplitude for this can written as
\begin{equation}
{\cal M}_{T,L} = -\kappa m_V^2 \left(1 - \frac{\partial {\rm Re\Pi_{T,L}}}{\partial \omega^2}\right)^{-1/2} \ ,
\end{equation}
where the factor $\left(1 - \partial {\rm Re\Pi_{T,L}}/\partial \omega^2\right)^{-1/2}$ comes from the wave function renormalization of the photon field due to the thermal correction. Therefore, the transition rate for a single transverse photon or longitudinal plasmon with four-momentum $k = (\omega,|\vec k|)$ can be written as
\begin{eqnarray}
\Gamma^{\rm trans}_{T,L} = \frac{1}{2\omega} \int \kappa^2 m_V^4 \left(1 - \frac{\partial {\rm Re\Pi_{T,L}}}{\partial \omega^2}\right)^{-1} d\Phi_1 \ ,
\end{eqnarray}
where $d\Phi_1 = (2\pi)^4\delta^4(k-p) \frac{d^3p}{2p^0(2\pi)^3}$ is the one particle phase space, and $p$ is the four-momentum of the outgoing dark photon satisfying the on shell condition $p^2 = m_V^2$. One finds,
\begin{equation}
\Gamma^{\rm trans}_{T,L} = \frac{ \pi \kappa^2 m_V^4}{\omega |1 - \partial{\rm Re\Pi_{T,L}}/\partial\omega^2|} \delta(\omega^2 - |\vec k|^2 - m_V^2) \ .
\end{equation}
Therefore, the production rate of dark photon can be written as
\begin{eqnarray}
\frac{d\Gamma_{T,L}^{{\rm prod},V}}{dV} &=& \int\frac{d^3k}{(2\pi)^3} \frac{1}{e^{\omega/T} - 1} \Gamma^{\rm trans}_{T,L} \nonumber \\
&=& \int \frac{|\vec k| d |\vec k|^2}{4\pi\omega} \frac{\kappa^2 m_V^4 \delta(\omega^2 - |\vec k|^2 - m_V^2)}{(e^{\omega/T} - 1)|1-\partial {\rm Re\Pi}_{T,L}/\partial\omega^2|}  \nonumber \\
&=& \int d\omega^2 \int \frac{|\vec k| d |\vec k|^2}{4\pi\omega} \delta(\omega^2 - |\vec k|^2 - {\rm Re\Pi}_{T,L})\frac{\kappa^2 m_V^4 \delta(\omega^2 - |\vec k|^2 - m_V^2)}{(e^{\omega/T} - 1)} \ ,
\end{eqnarray}
where the on shell conditions for the transverse photon and longitudinal plasmon are used in the last step. Upon integration over $|\vec k|^2$ we reproduce Eq.~(\ref{eq:gammares}).   

Finally, for a finite size thermal system, the radiation power into the dark photons can be written as
\begin{equation}
\label{power}
P = \int dV \int d\omega \left(\kappa_T^2 \frac{\omega d\Gamma^{\rm prod}_{T}}{d\omega dV} + \kappa_L^2 \frac{\omega d\Gamma^{\rm prod}_{L}}{d\omega dV}\right) \ ,
\end{equation}
In the next section, we use~(\ref{power}) to calculate the
$V$-emission rate as a function of~$m_V$.

\section{Solar luminosity and horizontal branch stars constraints on dark vectors}

\begin{figure}
\centering
\includegraphics[height=4in]{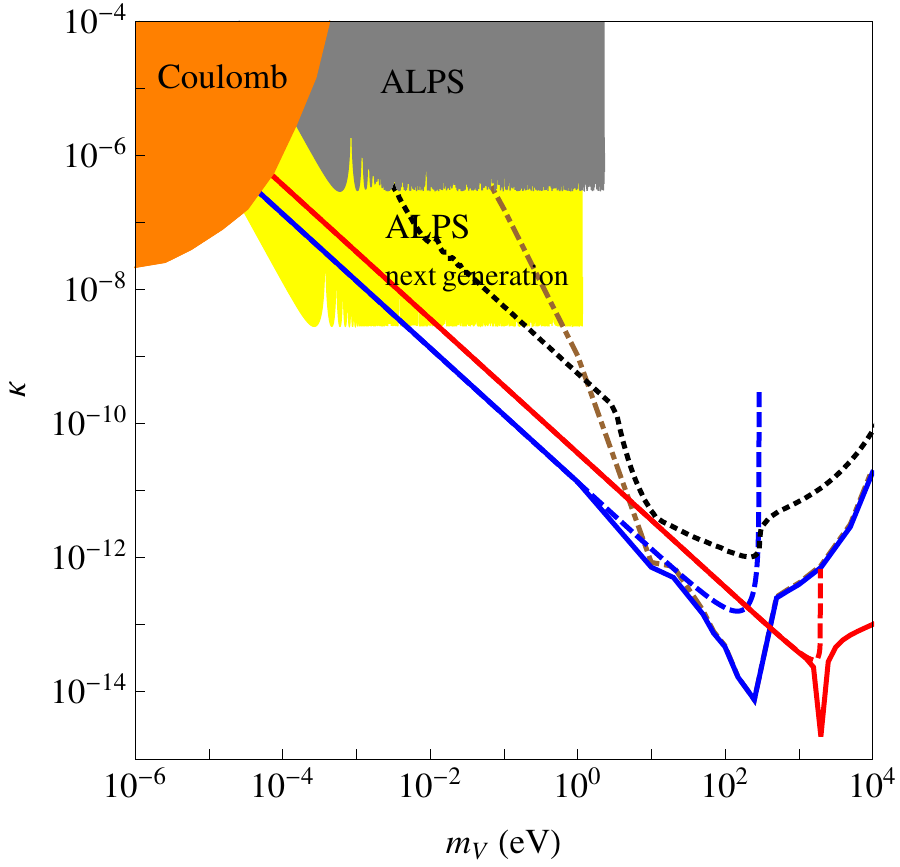}
\caption{Upper limits on the kinetic mixing parameter $\kappa$ vs.
  the mass of the dark photon $m_V$. The solid blue and red curves are
  the total constraints on the dark photon parameter space from the
  sun (blue) and horizontal branch stars (red).  The dashed curves show constraints
  from retaining only the longitudinal resonance contributions.  For
  comparison, the dot-dashed curve shows the upper limit by
  considering only the contribution from transverse modes and the dotted curve shows the constraint from the CAST experiment by only considering the contribution from the transverse mode~\cite{Redondo:2008aa}. The current
  bounds from the latest result of the LSW experiment by the ALPS
  collaboration~\cite{Ehret:2010mh} is shown in gray, accompanied by
  the potential reach (yellow region) of the next generation of LSW
  experiments~\cite{Redondo:2010dp}. }\label{fig:full}
\end{figure}

Inside the sun, the electrons are non-relativistic and
non-degenerate. In this limit, the leading order terms in a $T/m_e$
expansion of $\real \Pi_{T,L}$ can be written as
\begin{equation}
\real \Pi_{T} = \omega_{p}^{2} \ , \quad
\real\Pi_{L} = \omega_{p}^{2} \left(1 - \frac{|\vec k|^{2}}{\omega^{2}}\right) \ ,
\end{equation}
and the plasma frequency is given by
\begin{equation}
\omega_p^2 =\frac{e^2 n_e}{m_e} =\frac{4\pi \alpha n_e}{m_e} \ ,
\end{equation}
where $n_e$ is the number density of electrons. 

The production rate of the transverse modes is thoroughly calculated
in Ref.~\cite{Redondo:2008aa}, which we have checked and agree
with. As already mentioned multiple times, the production of
longitudinal modes is treated incorrectly in \cite{Redondo:2008aa}. We
believe that one can trace the error to $\Pi_L^{\rm Ref.\,\tiny
  \cite{Redondo:2008aa}}$ defined as $\omega_p^2-|\vec{k}|^2$ in
Eq.~(8) of Ref.~\cite{Redondo:2008aa}.  This is unphysical, as $\Pi_L$
generated by the plasma must vanish in the $n_e\to 0$ limit.  We can only speculate that this error
crept in from an inaccurate translation of the results from
Ref.~\cite{Braaten:1993jw} obtained in the Coulomb gauge to a generic
covariant gauge.

The production rate of the longitudinal modes has contributions both
from bremsstrahlung and from Compton scattering. Inside the Sun and a
horizontal branch stars the bremsstrahlung processes are dominant.  In the
non-relativistic and non-degenerate limit the cross section for the bremsstrahlung process to produce the
longitudinal massive photons at $k_\mu^2=m_V^2$ can be written as
\begin{equation}\label{eq:cx}
\left.\frac{d\sigma}{d \omega}\right|_{\rm brem} = \frac{\chi^2 Z^2 e^6 m_V^2 \sqrt{\omega^2 - m_V^2}}{ 3(2\pi)^3 m_e^2 \omega^4 v^2} \log\left|\frac{v + \sqrt{v^2 - 2 \omega/m_e}}{v - \sqrt{v^2 - 2 \omega/m_e}}\right| \ ,
\end{equation}
where $v$ is the velocity of the incoming electron. 
Then, appropriately
averaging over the electron energy distribution, the production rate 
({\em i.e.} the production rate of longitudinal dark vectors, when multiplied by the 
$\kappa^2_L$ factor)
can be written as
\begin{equation}\label{brem}
\left.\frac{d\Gamma^{\rm prod}_{L}}{dV d\omega}\right|_{\rm brem} = \sum_i\frac{8 Z_i^2 \alpha^3 n_e n_{Z_i}}{3 m_e^2} \frac{m_V^2}{\omega^4}\sqrt{\omega^2 - m_V^2} \sqrt{\frac{8 m_e}{\pi T}} f\left(\sqrt{\frac{\omega}{T}}\right) \ ,
\end{equation} 
where $n_{Z_i}$ is the number density of ions of charge $-Z_ie$, and 
\begin{equation}
f(a) = \int^\infty_a d x ~x e^{-x^2} \log\left|\frac{x+\sqrt{x^2-a^2}}{x-\sqrt{x^2-a^2}}\right| \ .
\end{equation}
In principle, inside the plasma, Eq.~(\ref{eq:cx}) should be modified to take into account the Debye screening effect characterized by the Debye screening length $\lambda_D^2 \equiv T/(e^2 n_e)$. However, inside the sun, a numerical study shows that the square of the typical momentum transfer of electrons $|\vec q|^2\sim m_e T$ is much larger than $1/\lambda_D^2$, and therefore this effect is of little importance for our level of rigour.

Substituting the explicit form of $\real \Pi_{T,L}$ into the resonant condition $m_V^2 = {\rm Re}\Pi_{T,L}$ discussed in Sec.~\ref{sec:3},
%\begin{equation}
%\kappa_L^2  = \frac{\kappa^2}{\left(1-\frac{\omega_p^2}{\omega^2}\right)^2 +  \left(\frac{{\rm Im}\Pi_{L}}{m_{V}^{2}}\right)^{2}}
%\end{equation}
we observe that the production rate of the transverse mode and longitudinal modes of dark
vectors reach a resonance at $m_V = \omega_p$ and $\omega = \omega_p$, respectively. Therefore, for
$m_V \ll \omega_p$ the longitudinal mode of dark photon can be produced on resonance at any
temperature. The resonant contribution to the radiation power per unit
volume and per unit frequency can be written as
\begin{equation}
\frac{d P_L}{d V d \omega} \approx \frac{1}{4\pi} \frac{\kappa^2 m_V^2 \omega_p^3}{e^{\omega_p/T} - 1} \times \delta(\omega-\omega_p)\ ,
\end{equation} 
which is independent of the details of the production processes. Therefore, the resonant contribution to power emitted into 
longitudinal dark vectors can be written as
\begin{equation}
\left.P_{\odot}\right|_{\rm res} \approx \kappa^2 m_V^2\int_0^{R_{\odot}} r^2 dr \frac{\omega_p^3(r)}{e^{\omega_p(r)/T} - 1} \ .
\end{equation}
where $R_{\odot}$ is the radius of the sun. The corresponding results in Ref. \cite{Redondo:2008aa} contain an additional 
factor of $2/\pi \times  m_V^2/\omega_p^2$ under the integral, which we argue is wrong. 

We use the standard solar model BP05(OP)~\cite{Bahcall:2004pz} to
calculate the total power radiated into dark photons. Consistency of
this model with observations, requires this additional "dark radiating
power" be smaller than the actual measured solar
luminosity~\cite{Frieman:1987ui,Raffelt:1988rx}, which is $L_\odot =
3.83\times10^{26}$ Watt.  This sets the limit on the parameter space
of the model, that we plot in Fig.~\ref{fig:full}.  The solid blue
curve shows the resulting constraint, and everything above this curve
is excluded.  For comparison, we also show the breakdown of the
constraint by different contributions.  The dashed blue curve shows
the constraint resulting from retaining only the resonant contribution
in the longitudinal modes. We can see that in the small $m_V$ region
$(m_V < 0.1{~\rm eV})$, the dark radiation is indeed dominated by the
longitudinal resonance contribution where the constraint can be
simplified as
\begin{equation}
\kappa \times \frac{m_V}{\rm eV} < 1.4\times10^{-11} \ .
\end{equation}

The dot-dashed brown curve shows the constraint by considering only
the contribution from the transverse mode, which coincides with the
solid blue curve in the large $m_V$ region, $m_V > 10$ eV, where we
are also in agreement with \cite{Redondo:2008aa}. However, in the
small $m_V$ region, the emission of $T$-modes gives only a subdominant
contribution, and indeed the solid blue curve positions itself much
below the brown dot-dashed curve. Therefore, our paper greatly
improves the bounds on dark photons below the mass range of a few~eV.

For the horizontal branch stars, we set the constraint by requiring 
that energy loss into dark photons should not exceed the nuclear energy 
generation rate. Thus,  the radiation power into dark photons in the nuclear reaction region of the horizontal branch stars should not exceed $10^{-5}$ Watt ${\rm gram}^{-1}$~\cite{Frieman:1987ui,Raffelt:1987yb}. The average temperature and density of the nuclear reaction region of horizontal branch stars can be estimated as
\begin{equation}
T_{\rm HB} = 10^8 K\ , \;\;\; \rho_{\rm HB} = 10^4 {\rm~gram/cm^3} \ .
\end{equation}
In this region, the electron gas can still be viewed as non-relativistic and non-degenerate~\cite{Grifols:1988fv}. The solid red curve in Fig.~\ref{fig:full} 
shows the upper limit on $\kappa$ by requiring the dark radiation power is smaller than the nuclear reaction power. The dashed red curve shows the upper limit by considering only the resonant longitudinal contribution. One can see that in the region $m_V < 100$ eV, the radiation is dominated by the longitudinal mode. The spike at $m_V\approx2$ keV is caused by the resonant production of the transverse modes which appears at $m_V^2 = \omega_p^2$. Overall, we find that the solar luminosity provides a somewhat stronger 
constraint for all masses below several 100 eV. 

\section{Summary and Discussions}

We have shown that the production rates of massive dark vectors $V$
with the St\"uckelberg mass $m_V$, coupled to the SM via the kinetic
mixing portal, scales as $m_V^2$ at small $m_V$ due to the emission of
the longitudinal modes of $V$.  This drastically change the strength
of the stellar constraints in the small $m_V$ region.  Thus, for the
first time, and despite the large abundance of literature on dark
photons, our paper sets correct stellar constraints on the dark photon
parameter space in the whole mass range below a few~eV.  This turns
out to be a region of special interest for LSW experiments.  In
Fig.~2, we show that even the most advanced ones
(ALPS)~\cite{Ehret:2010mh} find themselves inside a deeply excluded
region.  Recalling that the signal of dark photons in LSW experiments
scales as $\kappa^4$, a two order of magnitude gap in $\kappa$ at
$m_V\sim 0.01$ eV between stellar constraints and LSW region
translates into a required eight orders of magnitude improvement in
sensitivity before LSW experiments become competitive with stellar
bounds.  Indeed, a large part of the sensitivity region for the next
generation of LSW experiments, deemed reachable in
Ref.~\cite{Redondo:2010dp}, is also excluded.

We conclude with several additional remarks pertaining to light dark
photons:

\begin{itemize} 
\item The correct scaling with $m_V$, the $m_V^2$-behaviour, is
  important not only for the production, but also for the detection of
  dark photons in the laboratory environment.  Taking an example of
  dark matter detectors, made of some material with refractive index
  $n$, one can expect that in the region $m_V^2 \ll \omega^2 |1 - n|$,
  the absorption of the transverse mode scales as $m_V^4$, whereas the
  absorption of the longitudinal mode scales as $m_V^2$. Therefore, in
  this region one expect that the longitudinal mode will dominate the
  observed signal. The details of setting constraints on the solar
  dark photons with the use of the most sensitive low-energy threshold
  dark matter detectors will be addressed in our next paper
  \cite{ournextpaper}.

\item It has to be emphasized that in this paper we assume that, once
  produced, the dark photon freely escapes the stellar interior. It is
  possible that for large values of mixing angles, the energy loss
  process is quenched because of absorption. Determining whether
  such "islands" actually exist deep inside the excluded region, and
  if so whether they survive other constraints goes outside the scope
  of the present paper.

\item Stellar constraints derived in this paper have implications for
  the ``dark CMB''~\cite{Jaeckel:2008fi}.  We do not expect that the
  emission of the longitudinal modes in processes like $\gamma_T + e
  \to V_L + e$ in the primordial plasma will drastically change the
  estimates for the total energy density locked in $V$ modes, because
  of the very small value for $n_e/n_\gamma \simeq 6\times 10^{-10}$.
  The combination of our new stellar constraints derived in this
  paper, and the constraints from spectral distortions of the normal
  CMB in the very small $m_V$ region prevents generating large
  deviations of the effective number of relativistic degrees of
  freedom $N_{eff}$ from its standard model value.

\item Finally, we can go back to the picture of the ``Higgsed'' dark
  photon production, Eq. (\ref{rate_higgsed}), and ask the question of
  whether one can avoid strong constraints on $\kappa$ by choosing
  very small $\alpha'$, and by having the solar interior restore the
  symmetry in U(1)$_V$ sector, so that $m_V =0$ inside the sun. The
  symmetry restoration may occur for $\kappa \omega_p > m_V$, which
  can have an overlap with the LSW regions of interest. Such an
  unusual "chameleonic" scenario deserves a special analysis.

\end{itemize} 

{\bf Acknowledgements}  Research at the 
Perimeter Institute is supported in part by the Government of Canada through 
NSERC and by the Province of Ontario through MEDT.

\end{document}